\documentclass[aip,jap,reprint,amsmath,amssymb,showpacs,showkeys,superscriptaddress]{revtex4-1}

\usepackage{graphicx}
\usepackage{dcolumn}
\usepackage{bm}
\usepackage[mathlines]{lineno}
\linenumbers\relax 

\begin{document}


\title{Low-Dilution Limit of Zn$_{1\textrm{-}x}$Mn$_{x}$GeAs$_{2}$: Electrical and Magnetic Properties}

\author{L.~Kilanski}
 \email{kilan@ifpan.edu.pl}
\affiliation{Institute of Physics, Polish Academy of Sciences, Al. Lotnikow 32/46, 02-668 Warsaw, Poland}

\author{K.~Sza{\l}owski}
\affiliation{Department of Solid State Physics, Faculty of Physics and Applied Informatics, University of {\L}\'{o}d\'{z}, ul. Pomorska 149/153, 90-236 {\L}\'{o}d\'{z}, Poland}

\author{R.~Szymczak}
\author{M.~G\'{o}rska}
\author{E.~Dynowska}
\author{P.~Aleshkevych}
\author{A.~Podg\'{o}rni}
\author{A.~Avdonin}
\author{W.~Dobrowolski}
\affiliation{Institute of Physics, Polish Academy of Sciences, Al. Lotnikow 32/46, 02-668 Warsaw, Poland}

\author{I.~V.~Fedorchenko}
\author{S.~F.~Marenkin}
\affiliation{Kurnakov Institute of General and Inorganic Chemistry RAS, 119991 Moscow, Russia}

\date{\today}

\begin{abstract}

We present the studies of electrical transport and magnetic interactions in Zn$_{1\textrm{-}x}$Mn$_{x}$GeAs$_{2}$ crystals with low Mn content 0$\,$$\leq$$\,$$x$$\,$$\leq$$\,$0.042. We show that the ionic-acceptor defects are mainly responsible for the strong $p$-type conductivity of our samples. We found that the negative magnetoresistance (MR) with maximum values of about -50\% is related to the weak localization phenomena. The magnetic properties of Zn$_{1\textrm{-}x}$Mn$_{x}$GeAs$_{2}$ samples show that the random Mn-distribution in the cation sites of the host lattice occurs only for the sample with the lowest Mn-content, $x$$\,$$=$$\,$0.003. The samples with higher Mn-content show a high level of magnetic frustration. Nonzero Curie-Weiss temperature observed in all our samples indicates that weak ferromagnetic (for $x$$\,$$=$$\,$0.003) or antiferromagnetic (for $x$$\,$$>$$\,$0.005) interactions with the Curie-Weiss temperature, $|\Theta|$$\,$$<$$\,$3$\;$K, are present in this system. The Ruderman-Kittel-Kasuya-Yosida (RKKY) model, used to estimate the Mn-hole exchange integral $J_{pd}$ for the diluted Zn$_{0.997}$Mn$_{0.003}$GeAs$_{2}$ sample, makes possible to estimate the value of $J_{pd}$$\,$$=$$\,$(0.75$\pm$0.09)$\;$eV.

\end{abstract}





\maketitle

\section{Introduction}

Complex diluted magnetic semiconductors (DMS) are a subject of considerable interest in the recent years since they offer many  advantages over classical II-VI and III-V materials.\cite{Kossut93a, Dobrowolski03a} Ferromagnetic semiconductors with the Curie temperature, $T_{C}$, greater than room temperature, are needed for practical applications. However, most of the literature reports about the magnetic properties of DMS systems show the Curie temperatures much lower than 300$\;$K, which makes these compounds of little use for practical applications. The absence of room temperature ferromagnetic DMS systems creates the need for the development of new compounds fulfilling technological requirements. \\ \indent Recently II-IV-V$_{2}$ chalcopyrite semiconductors doped with transition metal ions have become a subject of considerable interest due to the appearance of room temperature ferromagnetism, creating the possibilities of utilizing them in spintronics.\cite{Erwin04a, Picozzi04a} It is known that the short-range magnetic interactions connected with the presence of magnetic clusters are responsible for high-temperature ferromagnetism in these alloys.\cite{Kilanski2009a, Kilanski2009b, Kilanski2010a} Among many experimental investigations showing ferromagnetism with large Curie temperatures due to short-range ordering there seems to be a shortage of studies devoted to low-dilution limits in which itinerant ferromagnetism might be induced. The possibility to control the magnetic properties of the material via changes in their electronic transport properties needed to fulfil the main aim of semiconductor spintronics,\cite{Ohno1998a} will be possible only when homogeneous materials will be technologically mastered and properly understood. \\ \indent In the present work we investigate magnetotransport and magnetic properties of Zn$_{1\textrm{-}x}$Mn$_{x}$GeAs$_{2}$ crystals with low Mn content $x$ varying in the range from 0 to 0.042. The present work extends our earlier research  devoted to both nanocomposite ZnGeAs$_{2}$:MnAs samples\cite{Kilanski2009a, Kilanski2009b, Kilanski2010a} and homogeneous Zn$_{1\textrm{-}x}$Mn$_{x}$GeAs$_{2}$ crystals.\cite{Kilanski2011a, Kilanski2012a} The low dilution Mn-alloying improves the structural quality of the samples allowing one to study long-range carrier mediated magnetic interactions in this system. Since Zn$_{1\textrm{-}x}$Mn$_{x}$GeAs$_{2}$ crystals are $p$-type semiconductors with large solubility of Mn-ions, the induction of the carrier-mediated-ferromagnetism seems to be a realistic aim. In order to understand the Mn-incorporation into this alloy the complexity of the magnetic interactions between Mn-ions needs to be understood.

\section{Basic characterization}

We present the studies of bulk Zn$_{1\textrm{-}x}$Mn$_{x}$GeAs$_{2}$ crystals grown using a direct fusion method from high purity ZnAs$_{2}$, Ge, and Mn powders taken in stoichiometric ratios.\cite{Novotortsev05a} The growth was performed at a temperature of about 1200$\;$K. Mn-doped crystals were cooled from the growth temperature down to 300$\,$K with a relatively high speed (about 5-10$\,$K/s) in order to improve the homogeneity of the samples and prevent Mn clustering.  \\ \indent The chemical composition of the samples was determined using the energy dispersive x-ray fluorescence method (EDXRF). A typical relative uncertainty of this method was about 10\%. The as-grown crystals were cut into slices with thickness of about 1.5$\;$mm prior to their structural characterization. The EDXRF analysis shows that our samples have Mn content $x$ changing in the range from 0 to 0.042. Moreover, it must be emphasized, that within our measurement accuracy all the studied crystals have the correct stoichiometry of Zn$_{1\textrm{-}x}$Mn$_{x}$:Ge:As equal to 1:1:2. \\ \indent The high resolution x-ray diffraction method (HRXRD) was used to investigate the structural properties of Zn$_{1\textrm{-}x}$Mn$_{x}$GeAs$_{2}$ crystals. Measurements were performed with the use of multipurpose X'Pert PRO MPD, Panalytical diffractometer with Cu K$_{\alpha1}$ radiation with wavelength $\lambda$$\,$=$\,$1.5406$\:$$\textrm{\AA}$, configured for Bragg-Brentano diffraction geometry and equipped with a strip detector and an incident-beam Johansson monochromator. In order to increase the accuracy and quality of the diffraction patterns the data acquisition in each individual measurement was done over several hours. The indexing procedure of measured diffraction patterns as well as calculations of the lattice parameters were performed using SCANIX 2.60PC program.\cite{Paszkowicz89a} \\ \indent The analysis of the HRXRD results allows us to identify two cubic disordered zincblende phases with lattice parameters, $a_{1}$ and $a_{2}$, equal to $a_{1}$$\,$$=$$\,$5.6462$\pm$0.0002$\,$$\textrm{\AA}$ and $a_{2}$$\,$$=$$\,$5.9055$\pm$0.0007$\,$$\textrm{\AA}$ for the pure ZnGeAs$_{2}$ crystal. The addition of a small quantity of Mn ($x$$\,$$=$$\,$0.003) to the Zn$_{1\textrm{-}x}$Mn$_{x}$GeAs$_{2}$ alloy results in stabilization of the tetragonal chalcopyrite structure with $a$$\,$$=$$\,$5.6751$\pm$0.0002$\,$$\textrm{\AA}$ and $c$$\,$$=$$\,$11.1534$\pm$0.0005$\,$$\textrm{\AA}$. Additionally, the cubic zincblende phase is observed in the Zn$_{1\textrm{-}x}$Mn$_{x}$GeAs$_{2}$ sample with $x$$\,$$=$$\,$0.003 with $a$$\,$$=$$\,$5.6471$\pm$0.0004$\,$$\textrm{\AA}$. Further increase of Mn content above $x$$\,$$=$$\,$0.003 results in a change of the main crystallographic phase of the alloy back to the cubic disordered zincblende structure. The lattice parameters determined for Zn$_{1\textrm{-}x}$Mn$_{x}$GeAs$_{2}$ crystals with $x$$\,$$>$$\,$0.01 are similar to those reported for the pure ZnGeAs$_{2}$ sample. The coexistence of disordered cubic zincblende and chalcopyrite tetragonal structures is justified by the phase diagram of the ternary Zn-Ge-As system\cite{Schon94a} in which both compounds lie on the same line connecting Ge and ZnAs$_{2}$. It must be emphasized, that diffraction patterns for both disordered zincblende and chalcopyrite structures (see Ref.~\onlinecite{Schon94a}) are located very close to each other and it was possible to distinguish between them only with the use of a state-of-the-art diffractometer. We want to emphasize that all our samples have almost the perfect stoichiometry of the ZnGeAs$_{2}$ compound, as determined with the use of the EDXRF technique. Therefore it is evident, that our alloy is based on ZnGeAs$_{2}$ compound, but the presence of the cubic disordered zincblende structure is a signature of a large chemical disorder of the alloy, widely observed in ternary chalcopyrite systems,\cite{Rincon92a} reflecting a mixing of the Zn and Ge atoms in the cation sublattice.

\section{Magnetotransport data}

In order to obtain information about fundamental electrical properties of the studied Zn$_{1\textrm{-}x}$Mn$_{x}$GeAs$_{2}$ alloy, temperature dependent magnetotransport measurements were performed. We have used the superconducting magnet with maximum magnetic field equal to $B$$\,$$=$$\,$13$\;$T and a sweep speed of about 0.5$\;$T/min, equipped with the cryostat allowing the control of the temperature of the sample in the range of 1.4$\,$$\leq$$\,$$T$$\,$$\leq$$\,$300$\;$K. The samples, cut to size of about 1$\times$1$\times$10$\;$mm, were etched and cleaned before making electrical contacts. The contacts were made with the use of gold wire and indium solder. The ohmic behavior of each contact pair was checked prior to proper measurements. The magnetoresistance and the Hall effect were measured simultaneously at selected temperatures.

\subsection{Basic magnetotransport characterization}

\begin{figure}[t]
  \begin{center}
   \includegraphics[width = 0.5\textwidth, bb = 110 130 715 500]{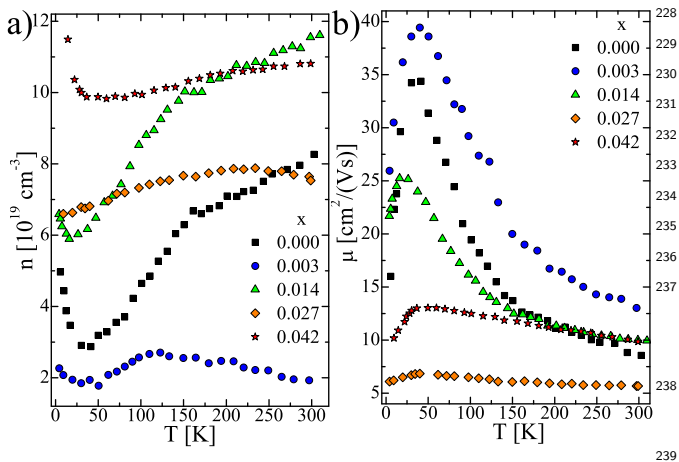}%
  \end{center}
   \caption{\label{FignmuvsT} The magnetotransport data including (a) Hall carrier concentration $n$ and (b) carrier mobility $\mu$ as a function of temperature for Zn$_{1\textrm{-}x}$Mn$_{x}$GeAs$_{2}$ samples containing different amount of Mn (see legend).}
\end{figure}

Initially, we measured the temperature dependence of the resistivity parallel to the current direction, $\rho_{xx}$, in the absence of an external magnetic field. Our results show, that in the case of all the samples metallic $\rho_{xx}$($T$) dependencies are observed, the behavior characteristic for degenerate semiconductors, i.e., an increase of the resistivity with an increasing temperature. This indicates that the carrier transport in the studied samples is not due to thermal activation of band carriers. It is evident, that defect states have major influence on the conductivity of this alloy. The characteristic feature observed for all investigated samples is the existence of a shallow minimum in $\rho_{xx}$($T$) dependence at temperatures below 50$\;$K. The existence of the minimum in $\rho_{xx}(T)$ dependence is likely to be related to the carrier scattering on the paramagnetic ions. Above $T$$\,$$=$$\,$50$\;$K the resistivity $\rho_{xx}$ is an increasing, nearly linear function of the temperature. \\ \indent The Hall effect measured as a function of temperature allows us to determine the temperature dependence of the Hall carrier concentration $n$ in all our samples (see Fig.$\;$\ref{FignmuvsT}a).
The results show that all our Zn$_{1\textrm{-}x}$Mn$_{x}$GeAs$_{2}$ crystals have a strong $p$-type conductivity with relatively high carrier concentration changing in the range of 10$^{19}$$\,$$\leq$$\,$$n$$\,$$\leq$$\,$10$^{20}$$\;$cm$^{-3}$. High concentration of conducting holes in this material is probably due to the existence of a large number of negatively charged Zn or Ge vacancy type defects and possibly other substitutional negatively charged defects. Inspection of results gathered in Fig.$\;$\ref{FignmuvsT}a shows that in the entire temperature range the Hall carrier concentration in the chalcopyrite Zn$_{0.997}$Mn$_{0.003}$GeAs$_{2}$ crystal is lower than in all disordered zincblende samples including ZnGeAs$_{2}$ with no Mn. Such a difference seems to be connected with thermodynamics of crystal growth which induced higher amount of point defects in the disordered zincblende structure crystal then that of chalcopyrite crystal. The higher concentration of electrically active defects resulted in an increase of the Hall carrier concentration. A general increase in the carrier concentration is observed in the studied disordered zincblende samples with an increase in the amount of Mn. The noticeable increase in the concentration of conducting holes indicates poor Mn allocation in the crystal lattice, which leads to the formation of vacancy type or interstitial point defects likely to be electrically active. \\ \indent The Hall carrier concentration $n$($T$) is a decreasing function of temperature at $T$$\,$$<$$\,$50$\;$K for most of the studied samples, while at $T$$\,$$>$$\,$50$\;$K the trend in the $n(T)$ dependence is opposite. Negative slope of $n(T)$ dependence at $T$$\,$$<$$\,$50$\;$K is a signature of a high ionic scattering mechanism involved in the carrier scattering at low temperatures. On the other hand, a positive slope of $n(T)$ dependence at $T$$\,$$>$$\,$50$\;$K is a signature that defect states responsible for metallic $\rho_{xx}(T)$ dependence are likely to be electrically active and being thermally activated are a source of conducting holes in our samples. \\ \indent The temperature dependent resistivity and the Hall effect data can be used to calculate the temperature dependence of the Hall mobility, $\mu$$\,$$=$$\,$($e$$\cdot$$n$$\cdot$$\rho_{xx}$)$^{-1}$, where $e$ is the elementary charge. The calculated $\mu$($T$) dependence for all our Zn$_{1\textrm{-}x}$Mn$_{x}$GeAs$_{2}$ samples is presented in Fig.$\;$\ref{FignmuvsT}b. In all samples the $\mu(T)$ dependence has a maximum at $T$$\,$$<$$\,$50$\;$K. Such a feature is characteristic for charged impurity scattering mechanism being important in the conduction of the material. At a maximum of the $\mu(T)$ dependence the carrier mobility reflects the superposition of a weak lattice and phonon scattering processes. At temperatures higher than 50$\;$K the $\mu(T)$ dependence is a decreasing function of temperature, which is a feature characteristic for phonon scattering mechanism. The highest value of the Hall carrier mobility, $\mu$$\,$$\approx$$\,$40$\;$cm$^{2}$/(V$\cdot$s), is observed in the chalcopyrite Zn$_{0.997}$Mn$_{0.003}$GeAs$_{2}$ sample (at $T$$\,$$\approx$$\,$50$\;$K) indicating, that the lattice scattering and the concentration of electrically active defects is for this particular sample the smallest in the entire series. A decrease in the maximum carrier mobility from 38$\;$cm$^{2}$/(V$\cdot$s) down to about 7$\;$cm$^{2}$/(V$\cdot$s) with the increase of Mn content in the samples is observed. It is a direct signature, that the allocation of Mn in the ZnGeAs$_{2}$ lattice was far from perfect and induced charged defects in the material.

\subsection{High Field Magnetotransport}

The isothermal magnetoresistance (MR) and the Hall effect measurements were performed for all our Zn$_{1\textrm{-}x}$Mn$_{x}$GeAs$_{2}$ samples. The $\rho_{xx}$($B$) curves were obtained by averaging the results for positive and negative current. For a simple data presentation the $\rho_{xx}$($B$) curves at different temperatures were normalized to the zero-field resistivity value $\rho_{0}$ by using the following relation: $\Delta \rho_{xx}/ \rho_{xx}(0)$$\,$$=$$\,$$(\rho_{xx}(B)$$\,$$-$$\,$$\rho_{xx}(B=0))$/$\rho_{xx}(B=0)$. In this manner, we calculated a number of isothermal magnetoresistance curves for each of the samples, obtained at various temperatures $T$$\,$$<$$\,$300$\;$K.
\begin{figure*}
    \includegraphics[width = 0.9\textwidth, bb = 0 40 803 560]
    {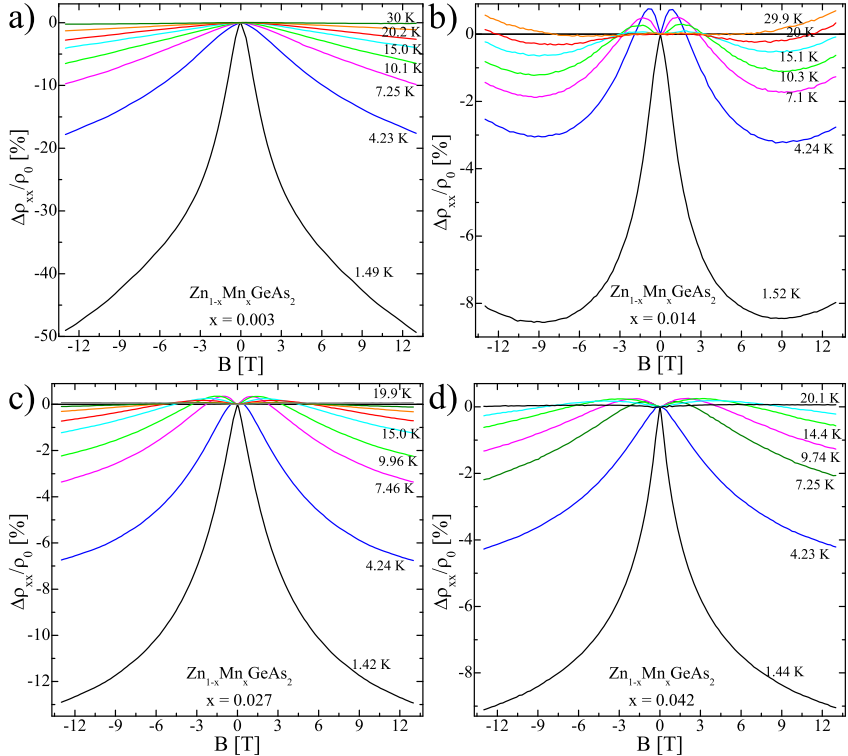}
  \caption{\small The magnetoresistance curves obtained experimentally at different temperatures for Zn$_{1\textrm{-}x}$Mn$_{x}$GeAs$_{2}$ samples with different chemical composition.}
  \label{FigMResCurves}
\end{figure*}
\\ \indent It should be noted, that MR for the nonmagnetic ZnGeAs$_{2}$ crystal is positive and is proportional to the square of the magnetic field for 1.4$\,$$<$$\,$$T$$\,$$<$$\,$300$\;$K. This effect can be clearly associated with the classical magnetoresistance due to the orbital motion of carriers in a magnetic field. Moreover, in the ZnGeAs$_{2}$ sample at low temperatures there are no other contributions to the MR, such as the weak localization of carriers on defect states.\cite{Anderson1958a}  \\ \indent The results of the magnetoresistance measurements for selected Zn$_{1\textrm{-}x}$Mn$_{x}$GeAs$_{2}$ samples performed at several stabilized temperatures are presented in Fig.$\;$\ref{FigMResCurves}. The results presented in Fig.$\;$\ref{FigMResCurves} indicate that the addition of a small quantity of Mn to the Zn$_{1\textrm{-}x}$Mn$_{x}$GeAs$_{2}$ alloy results in a drastic change of the MR behavior with respect to the nonmagnetic ZnGeAs$_{2}$ crystal. The results show a negative MR for our Zn$_{1\textrm{-}x}$Mn$_{x}$GeAs$_{2}$ samples, present at temperatures lower than 30$\;$K. Magnetoresistance in Zn$_{1\textrm{-}x}$Mn$_{x}$GeAs$_{2}$ samples with $x$$\,$$\leq$$\,$0.042, studied in this paper have a different shape and an amplitude of the order of magnitude higher than that studied in Ref.$\;$\onlinecite{Kilanski2011a} (Zn$_{0.947}$Mn$_{0.053}$GeAs$_{2}$ sample of similar concentration and mobility of carriers at $T$$\,$$=$$\,$1.4$\;$K). It is therefore clear that in the case of currently studied samples the magnetoresistance should be associated with a different mechanism than the spin-disorder scattering process.\cite{Kilanski2011a, Gennes1958a} \\ \indent The addition of a small quantity ($x$$\,$$=$$\,$0.003) of Mn to the Zn$_{1\textrm{-}x}$Mn$_{x}$GeAs$_{2}$ alloy results in a drastic change of the MR with respect to the nonmagnetic ZnGeAs$_{2}$ sample. The chalcopyrite Zn$_{1\textrm{-}x}$Mn$_{x}$GeAs$_{2}$ sample with $x$$\,$$=$$\,$0.003 shows negative MR with maximum value of about -50\% at $T$$\,$$=$$\,$1.42$\;$K, slowly decreasing with the temperature to about 0.2\% at $T$$\,$$=$$\,$20$\;$K. The MR at $T$$\,$$>$$\,$30$\;$K does not show any signatures of negative contribution and we can observe only a small positive contribution to the MR (with values less than 1\%), proportional to the square of the magnetic field and slowly decreasing as a function of temperature. The positive MR at $T$$\,$$>$$\,$30$\;$K is caused by the orbital motion of conducting carriers in the presence of the external magnetic field. \\ \indent The MR curves for our disordered zincblende Zn$_{1\textrm{-}x}$Mn$_{x}$GeAs$_{2}$ samples with $x$$\,$$\geq$$\,$0.014 show much smaller amplitudes (almost an order of magnitude lower) with respect to the chalcopyrite Zn$_{0.997}$Mn$_{0.003}$GeAs$_{2}$ sample. The maximum amplitude of negative MR equals 13\% for the sample with $x$$\,$$=$$\,$0.027 at $T$$\,$$\approx$$\,$1.4$\;$K. Moreover, at low magnetic fields and at temperatures higher than 2$\;$K we observe a positive contribution to the MR scaling with the square of the magnetic field, caused by the orbital MR. \\ \indent We performed the scaling analysis of the negative MR observed in our samples at temperatures lower than 30$\;$K. The scaling analysis to $\Delta \rho_{xx}$/$\rho_{0}$$\,$$\propto$$\,$$B^{m}$ proportionality, where $m$ is the scaling factor, shows, that the MR results for all our samples can be fitted with a good accuracy with the values of scaling factor around 0.6$\,$$\leq$$\,$$m$$\,$$\leq$$\,$0.8. The value of the exponent $m$ carries information about the possible physical mechanism of the negative MR. The Moriya-Kawabata spin-fluctuation theories\cite{Moriya1973a, Moriya1973b} predict that the negative magnetoresistance should scale with $m$$\,$$=$$\,$1 and 2 for weakly and nearly ferromagnetic metals, respectively. Since the $m$ values for our samples are lower than the above values we can exclude the spin-fluctuations from being the main physical mechanism responsible for the observed negative MR. Moreover, the dependence of the MR on the magnetization $M$ normalized to the saturation magnetization $M_{S}$ ($M$/$M_{S}$) is neither square nor cubic. It is another signature that the negative MR for our Zn$_{1\textrm{-}x}$Mn$_{x}$GeAs$_{2}$ samples originates from processes that are not directly related with magnetic impurity scattering. \\ \indent Theories of quantum corrections to the conductivity predict that weak localization (WL) phenomena can be responsible for negative MR present in many different disordered metals and semiconductors at low temperatures.\cite{Lee1985a} Localization shortens the mean free path of the carriers and gives rise to a mobility edge in the valence band. The product $k_{F}l$, where $k_{F}$ is the quasi-Fermi wave vector and $l$ is the mean free path, reflects the degree of the disorder and localization in a material. For $k_{F}l$$\,$$\sim$$\,$1 the disorder and electron localization are strong and the conducting holes cannot be thermally activated to the valence band, where they could diffuse thorough the crystal. The magnetic field and the presence of magnetic impurities destroys carrier localization allowing carriers to diffuse. Experimentally, it is visible as negative MR. The negative MR, caused by the WL in the presence of spin-orbit interaction and magnetic impurity scattering is predicted to show complex behavior. A number of different corrections have to be considered to fit the experimental MR results.\cite{Baxter1989a} The $k_{F}l$ factor at $T$$\,$$<$$\,$30$\;$K for all our samples indicates that the localization is strong. The theories of MR due to WL predict a presence of both positive (at low magnetic fields) and negative MR (at higher fields) for low diffusivity systems (0.5 cm$^{2}$/s). Moreover, the MR in our samples decreases with increasing the average Mn content, $x$, and the related effective Mn content $\bar{x_{Mn}}$ derived from susceptibility and magnetization data. Thus, the observed MR is not related to a magnetic impurity scattering mechanism that would lead to MR proportional to the amount of magnetic impurities in the crystal. WL is known to be destroyed by the presence of magnetic impurities in the material. Therefore, the destruction of the WL by the presence of an increasing number of magnetic impurities in the crystals explains the observed MR in the Zn$_{1\textrm{-}x}$Mn$_{x}$GeAs$_{2}$ samples with $x$$\,$$>$$\,$0.003.

\section{Magnetic properties}

\noindent Magnetic properties of the Zn$_{1\textrm{-}x}$Mn$_{x}$GeAs$_{2}$ samples are studied by means of both ac and dc magnetometry. The mutual inductance method, employed into the LakeShore 7229 susceptometer/magnetometer system was used in order to determine the temperature dependencies of the ac magnetic susceptibility. The high field magnetization was measured with the use of a Quantum Design XL-5 Magnetometer.

\subsection{Low field results}

\begin{figure}[t]
 \includegraphics[width = 0.42\textwidth, bb = 0 20 590 530]
 {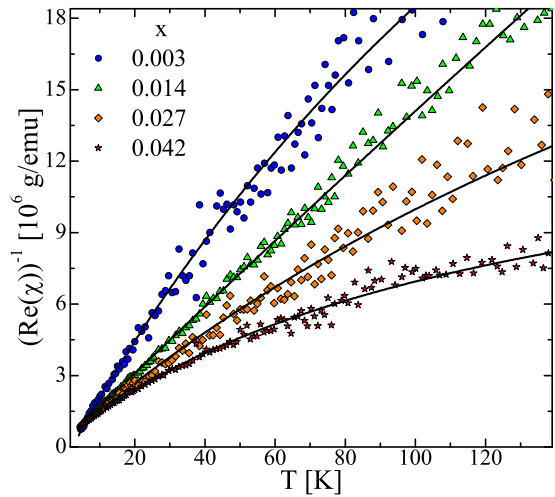}%
 \caption{\label{FigReXvsT} The inverse of the magnetic susceptibility as a function of temperature for Zn$_{1\textrm{-}x}$Mn$_{x}$GeAs$_{2}$ samples containing different amount of Mn (see legend).}
\end{figure}

\begin{figure}[t]
 \includegraphics[width = 0.42\textwidth, bb = 0 20 590 530]
 {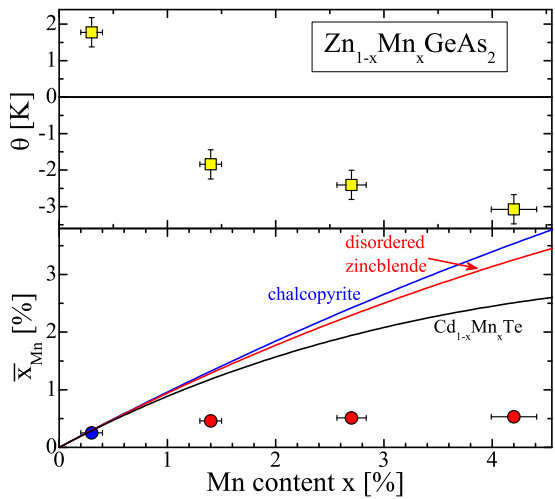}%
 \caption{\label{FigThetaCvsx} The Curie-Weiss temperature and the effective magnetically-active Mn-content as a function of the chemical composition for the Zn$_{1\textrm{-}x}$Mn$_{x}$GeAs$_{2}$ samples. Markers - experimental data, solid lines calculated for random distribution of magnetic ions (see text).}
\end{figure}

The temperature dependence of the ac magnetic susceptibility was measured in the temperature range 4.3$\,$$\leq$$\,$$T$$\,$$\leq$$\,$180$\;$K. During the measurement the sample was put into the alternating magnetic field having amplitude $H_{AC}$$\,$$=$$\,$10$\;$Oe and frequency $f$$\,$$=$$\,$625$\;$Hz. The results in the form of the temperature dependencies of the inverse of the real part of the magnetic susceptibility (Re($\chi_{AC}$))$^{-1}$($T$) obtained in the case of several Zn$_{1\textrm{-}x}$Mn$_{x}$GeAs$_{2}$ samples having different Mn-content are shown in Fig.$\;$\ref{FigReXvsT}. As we can see all the samples showed paramagnetic Curie-Weiss behavior of the (Re($\chi_{AC}$))$^{-1}$($T$) at temperatures between 4.3$\,$$\leq$$\,$$T$$\,$$\leq$$\,$30$\;$K. In II-IV-V$_{2}$ DMSs, we can distinguish two major components determining their magnetic properties. The first is the paramagnetic component introduced by Mn$^{2+}$ ions with half filled 3d$^{5}$ shell in the ground state with $S$$\,$$=$$\,$5/2. The second term is the diamagnetic one originating from the nonmagnetic host lattice and substitutional diamagnetic ions. However, at higher temperatures, one can see the deviation of the (Re($\chi_{AC}$))$^{-1}$($T$) from a paramagnetic temperature dependence towards lower values. It indicates the existence of an additional, temperature independent term in the (Re($\chi_{AC}$))$^{-1}$($T$) curves.
The Van-Vleck paramagnetism can be a source of the additional term in the magnetic susceptibility. Such an effect with the susceptibility values around 10$^{-7}$$\;$emu/g was observed in Cd$_{1\textrm{-}x}$Co$_{x}$Se.\cite{Lewicki1990a} However, in our samples the additional susceptibility is about an order of magnitude higher. The additional susceptibility in our samples can be associated with the presence of magnetic ions coupled by the short range magnetic interactions. \\ \indent The temperature dependence of the inverse of the magnetic susceptibility (Re($\chi_{AC}$))$^{-1}$($T$) for the temperatures well above the Curie-Weiss temperature can be fitted with the use of the modified Curie-Weiss law in the form:
\begin{eqnarray}
   \chi(T) &=& \frac{C}{T - \Theta} + \chi_{dia} + \chi_{p}, \; \textrm{with}  \label{EqCWLaw} \\
   C&=& \frac{N_{0} g^{2} \mu_{B}^{2} S(S+1) {\bar x_{Mn}}}{3k_{B}} \label{EqCWLawCConst}
\end{eqnarray}
where $C$ is the Curie constant, $\Theta$ is the paramagnetic Curie-Weiss temperature, $\chi_{dia}$$\,$$=$$\,$$-2\times$10$^{-7}$$\;$emu/g is the diamagnetic contribution to the magnetic susceptibility originating from the host lattice (The value was determined from our magnetization measurements of ZnGeAs$_{2}$), $\chi_{p}$ is the paramagnetic contribution to the magnetic susceptibility originating from short-range coupled Mn-ions, $N_{0}$ is the number of cation sites per gram, $g$$\,$$\simeq$$\,$2 is the effective spin-splitting factor, $S$$\,$$=$$\,$5/2 is the spin-magnetic momentum of the Mn ions, $\mu_{B}$ is the Bohr magneton, $k_{B}$ is the Boltzmann constant, and ${\bar x_{Mn}}$ is the effective magnetically-active Mn content. The experimental (Re($\chi_{AC}$))$^{-1}$($T$) curves for the temperatures higher than 20$\;$K are fitted to Eq.\ref{EqCWLaw} (see Fig.$\;$\ref{FigReXvsT}) with the values of $C$, $\Theta$, and $\chi_{p}$ as fitting parameters. The contribution to the magnetic susceptibility originating from the short-range-coupled Mn ions is found to be an an increasing function of the Mn-content changing from the value of 3$\times$10$^{-7}$$\;$emu/g up to 1.3$\times$10$^{-6}$$\;$emu/g. It is a signature that the amount of paramagnetic ion pairs forming Mn-As-Mn configurations coupled with superexchange magnetic interaction increases with increasing concentration of Mn in the sample. \\ \indent The fitting procedure shows that both $\Theta$ and $C$ are highly composition dependent. The values of the Curie constant are used to calculate the magnetically-active Mn-content, ${\bar x_{Mn}}$, using Eq.$\;$\ref{EqCWLawCConst}. As a result of the fitting procedure the values of the Curie-Weiss temperature and the amount of magnetically active Mn ions are evaluated. The results of the fitting procedure are gathered in Fig.$\;$\ref{FigThetaCvsx}.
The results show that the Curie-Weiss temperature changed sign with an increase of the Mn content. This indicates the change of the dominant magnetic interactions in this system from ferromagnetic at low composition $x$ to antiferromagnetic for $x$ between 0.003 and 0.014. It is therefore possible that, with the increase of Mn in the alloy above $x$$\,$$=$$\,$0.003, the amount of Mn not diluted randomly in the crystal increases. At the same time we observe that the amount of magnetically active Mn ions, ${\bar x_{Mn}}$, is close to the average Mn content, $x$, only in the case of a very diluted Zn$_{0.997}$Mn$_{0.003}$GeAs$_{2}$ sample. In addition to the above mentioned sample, for all our crystals with 0.014$\,$$\leq$$\,$$x$$\,$$\leq$$\,$0.042 the quantity of magnetically active Mn ions remaining in the state of high-spin $J$$\,$$=$$\,$$S$$\,$$=$$\,$5/2 does not exceed $\bar{x}_{Mn}$$\,$$\approx$$\,$0.005$\pm$0.001. It is evident, that the majority of Mn ions in Zn$_{1\textrm{-}x}$Mn$_{x}$GeAs$_{2}$ samples with $x$$\,$$\geq$$\,$0.014 do not substitute for Zn sites in the crystal lattice and therefore possess smaller net magnetic moment than Mn$^{2+}$. Moreover, it is also probable that a large fraction of Mn ions in these samples occupies  interstitial sites of the crystal lattice, which promotes short range superexchange interactions leading to antiferromagnetic pairing of Mn ions and zero net magnetic moment of such pairs. The antiferromagnetic state of Mn ions with no net magnetic moment is energetically preferred for Zn$_{1\textrm{-}x}$Mn$_{x}$GeAs$_{2}$ system with $x$$\,$$=$$\,$0.25 and 0.50 [\onlinecite{Zhao2012a}]. It is therefore highly probable that the magnetic ions form antiferromagnetic states in this semiconductor matrix when the $x$ value is higher than 0.25. \\ \indent The data allow us to comment on the distribution of magnetic Mn ions in the sample. In the presence of short-range antiferromagnetic interactions a fraction of Mn ions is magnetically inactive, i.e. involved in clusters showing zero or low ground state spin (dominantly antiferromagnetically coupled pairs for low $x$). Therefore, the main contribution to magnetization originates from ions which do not possess nearest-neighbors in the sublattice occupied by magnetic ions and thus do not belong to clusters. If the distribution of ions within the magnetic sublattice is purely random, the amount of magnetically active ions can be estimated as ${\bar x}=x\left(1-x\right)^{z}$, where $z$ is the maximum number of nearest-neighbor magnetic ions which can couple antiferromagnetically to a given ion. For typical diluted magnetic semiconductors having a zincblende structure, the corresponding formula is ${\bar x}=x\left(1-x\right)^{12}$ (see Ref.$\;$\onlinecite{Behringer1958a}). In the case of our samples, two crystalline structures were detected, as mentioned in the section II. We assume that Mn ions couple antiferromagnetically through the superexchange mechanism via As ions. Therefore, for the chalcopyrite structure, we have ${\bar x}=x\left(1-x\right)^{4}$. On the other hand, for a quaternary sample with a  disordered zincblende structure Zn$_{1\textrm{-}x}$Mn$_{x}$GeAs$_{2}$ the Mn content $x$ applies only to half of the cations, therefore we should calculate ${\bar x}=x\left[1-\left(x/2\right)\right]^{12}$. In Fig. 4 we plot by solid lines the estimates of active Mn content for both crystalline structures found in our samples, as well as an estimate for a classical diluted magnetic semiconductor, Cd$_{1\textrm{-}x}$Mn$_{x}$Te. It is apparent that for $x$$\,$$>$$\,$0.003, the calculated values of active Mn concentration are several times higher than those found experimentally. Such a discrepancy is a signature of the fact that Mn ions are not distributed randomly in our samples.

\subsection{High field results}

\begin{figure}[t]
 \includegraphics[width = 0.42\textwidth, bb = 0 20 590 530]
 {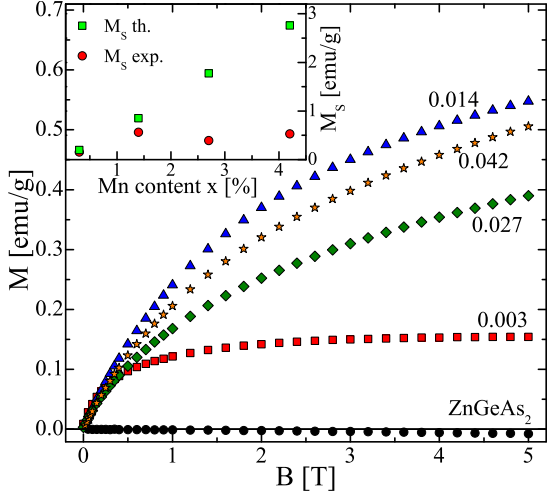}%
 \caption{\label{FigMvsT} The isothermal magnetic field dependencies of the magnetization measured at $T$$\,$$=$$\,$2$\;$K for our Zn$_{1\textrm{-}x}$Mn$_{x}$GeAs$_{2}$ samples. The inset shows the magnetization $M_{S}$ exp. obtained from fitting our experimental data with Eqs.$\;$\ref{Eq03} and \ref{Eq03a} (circles) and the saturation magnetization $M_{S}$ th. obtained from Eq.$\;$\ref{Eq03a} by assuming $\bar{x}$$\,$$=$$\,$$x$ (rectangles).}
\end{figure}

The magnetization of our samples was studied with the use of the Quantum-Design Superconductive-Quantum-Interference-Device (SQUID) magnetometer system. The SQUID magnetometer enabled precise measurements of the magnetic moment of the samples as a function of the magnetic field up to $B$$\,$$=$$\,$5$\;$T and at temperatures in the range 2$\,$$\leq$$\,$$T$$\,$$\leq$$\,$100$\;$K. The isothermal magnetic field dependence of the magnetization was measured for the samples with different chemical content. The experimental results were corrected by subtracting the contribution of the sample holder. Magnetization of all the samples does not show magnetic hysteresis for temperatures above 2$\;$K. That may indicate a lack of macroscopic MnAs precipitates and that the long-range RKKY itinerant interaction does not produce magnetic order at $T$$\,$$>$$\,$2$\;$K. Examples of $M$($B$) curves obtained for the selected Zn$_{1\textrm{-}x}$Mn$_{x}$GeAs$_{2}$ samples with different chemical composition are presented in Fig.$\;$\ref{FigMvsT}.
The magnetization shows a diamagnetic response for the nonmagnetic ZnGeAs$_{2}$ sample indicating the absence of non-intentional paramagnetic ion doping present in our pure sample, in agreement with the previous result based on the susceptibility results. The magnetization curve observed for the chalcopyrite Zn$_{1\textrm{-}x}$Mn$_{x}$GeAs$_{2}$ sample with $x$$\,$$=$$\,$0.003 shows a behavior characteristic of a paramagnet, i.e., the $M$($B$) curve can be easily fitted with the use of Brillouin function. Moreover, the saturation of the $M$($B$) curve for the sample with $x$$\,$$=$$\,$0.003 is reached at the magnetic field $B$$\,$$\approx$$\,$3$\;$T. It is a signature of a random Mn-distribution in the host lattice, and probably lack of significant antiferromagnetic Mn-pairing. The addition of a higher quantity of Mn to the alloy resulted in a different shape of the $M$($B$) curve for our crystals with $x$$\,$$>$$\,$0.01. As we can see in Fig.$\;$\ref{FigMvsT} the magnetization does not reach saturation even at $B$$\,$$=$$\,$5$\;$T for $x$$\,$$>$$\,$0.003. \\ \indent We fitted our experimental data to the expression\cite{Gaj1979a}
\begin{equation}\label{Eq03}
    M = M_{S} B_{J}\Bigg{(}\frac{g \mu_{B} J B}{k_{B} (T + T_{0})}\Bigg{)}+ \chi_{dia} B,
\end{equation}
where
\begin{equation}\label{Eq03a}
    M_{S} = \bar{x} N_{0} \mu_{B} g J,
\end{equation}
and $B_{J}$ is the Brillouin function. The term $\chi_{dia} B$ represents the diamagnetic contribution of the ZnGeAs$_{2}$, $g$ is the $g$-factor of the magnetic ion (for Mn $g$$\,$$=$$\,$2), $\mu_{B}$ is the Bohr magneton, $J$$\,$$=$$\,$$S$$\,$$=$$\,$5/2 is the total magnetic momentum of the Mn$^{2+}$ ion, $k_{B}$ is the Boltzmann constant, $T$ is the temperature, and $N_{0}$ is the number of cation sites per gram. \\ \indent The two fitting parameters, $\bar{x}$ and $T_{0}$, represent the amount of magnetically active Mn ions in the material and the exchange interaction among magnetic ions, respectively. The $T_{0}$ values obtained for our samples were close to the estimated Curie-Weiss temperatures, $\Theta$, and decreased as a function of $x$: for $x$$\,$$=$$\,$0.003 $T_{0}$$\,$$=$$\,$1.13$\;$K, for $x$$\,$$=$$\,$0.014 $T_{0}$$\,$$=$$\,$-1.47$\;$K, for $x$$\,$$=$$\,$0.027 $T_{0}$$\,$$=$$\,$-1.50$\;$K, and for $x$$\,$$=$$\,$0.042 $T_{0}$$\,$$=$$\,$-1.95$\;$K. The estimated error in $T_{0}$ is about 20\%. \\ \indent In the inset to Fig.$\;$\ref{FigMvsT} we show the magnetization $M_{S}$ th. calculated using Eq.$\;$\ref{Eq03a} with $\bar{x}$$\,$$=$$\,$$x$ and the magnetization $M_{S}$ exp. obtained from fitting our experimental data to Eqs.$\;$\ref{Eq03} and \ref{Eq03a} with $\bar{x}$ and $T_{0}$ as fitting parameters. \\ \indent We used the Eqs.$\;$\ref{Eq03} and \ref{Eq03a} to estimate the amount of magnetically active Mn ions in our samples, $\bar{x}$. The calculated $x$/$\bar{x}$ ratios are equal to 1.28 and 1.51 for $x$$\,$$=$$\,$0.003 and 0.014, respectively, and increases dramatically up to 4.5 and 5.2 for $x$$\,$$=$$\,$0.027 and 0.042, respectively. It indicates rather poor Mn allocation in the Zn$_{1\textrm{-}x}$Mn$_{x}$GeAs$_{2}$ semiconductor matrix for $x$$\,$$\geq$$\,$0.014 and that the level of frustration in the material increases as a function of $x$. The $M$($B$) results and their interpretation are in agreement with the susceptibility data.

\subsection{Electron paramagnetic resonance}

\begin{figure}[t]
 \includegraphics[width = 0.42\textwidth, bb = 0 20 590 530]
 {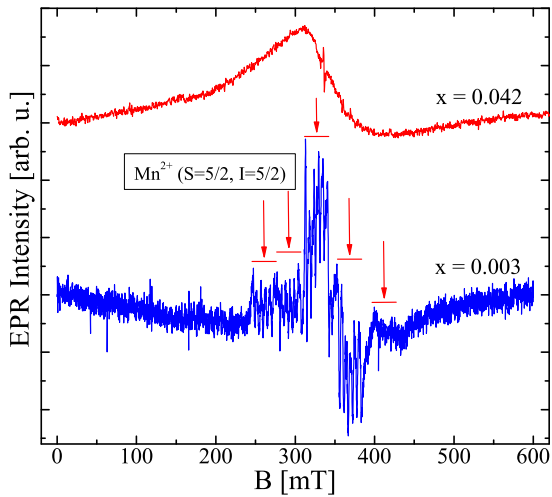}%
 \caption{\label{FigEPR} The EPR spectra measured at $T$$\,$$=$$\,$10$\;$K for selected Zn$_{1\textrm{-}x}$Mn$_{x}$GeAs$_{2}$ samples with different chemical composition.}
\end{figure}

The electron paramagnetic resonance (EPR) spectra were measured with the use of a Bruker EMX spectrometer. The sample was put into a microwave radiation system with fixed frequency 9.40$\;$GHz. The temperature of the sample was controlled with the use of Oxford Instruments helium flow cryostat. The measurements performed at selected temperatures showed a signal related to the presence of Mn ions only at low temperatures around 10$\;$K. The normalized EPR spectra obtained for two selected Zn$_{1\textrm{-}x}$Mn$_{x}$GeAs$_{2}$ samples are gathered in Fig.$\;$\ref{FigEPR}.
Since our samples are polycrystalline the EPR results are the average over all possible directions of crystallites with respect to the external magnetic field axis. The EPR results for low Mn content Zn$_{0.997}$Mn$_{0.003}$GeAs$_{2}$ indicate the presence of thirty resonance lines - a pattern, characteristic of Mn$^{2+}$ ions showing both fine and hyperfine structure components, due to electron spin $S$$\,$$=$$\,$5/2 and nuclear spin $I$$\,$$=$$\,$5/2, respectively. We believe that five groups, each consisting of six lines belonging to Mn$^{2+}$, are observed (marked in Fig.$\;$\ref{FigEPR}). The shift between the five line groups is due to the presence of zero-field-splitting of the ground state levels. Clearly, the EPR results show the 2+ charge state of Mn ions in the Zn$_{0.997}$Mn$_{0.003}$GeAs$_{2}$ sample. This interpretation would be consistent with the magnetometery data. For the Zn$_{1\textrm{-}x}$Mn$_{x}$GeAs$_{2}$ with Mn content higher than $x$$\,$$=$$\,$0.003, due to the broadening of the resonance lines, both the fine and hyperfine structure of Mn$^{2+}$ becomes unresolved. The value of the electron effective $g$ factor for our samples, as estimated from the EPR spectrum, equals 1.9972.

\subsection{Curie-Weiss temperature and determination of $J_{pd}$}

Let us discuss the dependence of the Curie-Weiss temperature on the magnetic component concentration $x$ for a diluted magnetic semiconductor. If both short-range antiferromagnetic couplings and long-range net ferromagnetic interactions are present in the system, the value of $\Theta$ can be written as\cite{Ferrand01} $\Theta={\bar x}\Theta_{F}+\Theta_{AF}$, where
\begin{equation}\label{Eq04}
\Theta_{F}=\frac{S(S+1)}{3k_{\rm B}}\sum_{k}^{}{z_{k}J\left(r_{k}\right)},
\end{equation}
and $\Theta_{AF}$ is an antiferromagnetic term, depending on $x$ but not directly proportional, $z_k$ is the number of lattice sites at a distance of $r_{k}$ from the selected site at the origin, and $J(r_{k})$ is the exchange interaction between magnetic ions separated by a distance $r_{k}$. Let us emphasize that the summation is performed only over the lattice sites which can be occupied by substitutional magnetic ions.  We exclude the ions involved in nearest-neighbour pairs from the sum in Eq.~\ref{Eq04}, since we assume that they are blocked by strong short-range antiferromagnetic couplings. Using the effective concentration, ${\bar x}$, determined from fitting the experimental data to Eqs.$\;$\ref{Eq03} and \ref{Eq03a} is based on the assumption that only this fraction of magnetic ions is magnetically active, at least in the range of applied magnetic fields and temperatures used.

For the case of the sample with the lowest Mn content ($x=0.003$), for which the Curie-Weiss temperature is positive (ferromagnetic), we made an attempt to explain the Curie-Weiss temperature value by applying a Ruderman-Kittel-Kasuya-Yosida (RKKY) model of long-range magnetic interactions.\cite{Ruderman54,Kasuya56,Yosida57} For that purpose we assume that the antiferromagnetic contribution, $\Theta_{AF}$, is negligible due to an ultralow concentration of magnetic ions, and that only a ferromagnetic contribution to Curie-Weiss temperature exists. In our considerations, the appropriate distances and numbers of lattice sites at a given distance are generated for a chalcopiryte lattice (in which Mn ions substitute for Zn ions) with lattice constants $a$ and $c$ discussed in the Section II.

The interaction between the magnetic ions is assumed to be the RKKY interaction for disordered three-dimensional system \cite{Ruderman54,Kasuya56,Yosida57,deGennes62}:
\begin{align}\label{Eq05}
J\left(r\right)=&N_{V}\frac{J_{pd}^2 m^{*}a^4c^2}{8\pi^3 \hbar^2}\,k_{\rm F}^4 \nonumber \\
&\times\,\frac{\sin\left(2k_{\rm F} r\right)-\left(2k_{\rm F} r\right)\cos\left(2k_{\rm F} r\right)}{\left(2k_{\rm F} r\right)^4}\,\exp\left(-r/\lambda\right),
\end{align}
where $k_{\rm F}=\left(3\pi^2 n/N_{V}\right)$ is the Fermi wavevector and $J_{pd}$ is the Mn-hole exchange interaction constant.

In the calculations, we assumed the effective mass of the charge carriers $m^{*}/m_{e}=0.4$ (similar to that in GaAs) and the number of valleys $N_{V}=3$. The characteristic decay length, $\lambda$, can be identified as a mean free path for the charge carriers. Such a quantity can be estimated using Drude model of conductivity from the formula $\lambda=\hbar k_{\rm F}\mu/e$, where $\mu$ is the mobility of the charge carriers and $e$ denotes elementary charge value. In order to determine the $\lambda$ value we used the values of charge carrier concentration and mobility for the sample with $x=0.003$ at the lowest available temperature, i.e. about 4 K, which yielded the result of $\lambda=10$ \AA. For the accepted parameter values, the experimental value of $\Theta=\left(1.8\pm 0.4\right)$ K would yield $J_{pd}=\left(0.75\pm 0.09\right)$ eV. Let us note that the given uncertainty in the $J_{pd}$ value is only due to the corresponding uncertainties in the Curie-Weiss temperature and effective concentration of magnetic ions. Additionally, we mention that one of the factors influencing the determined value of exchange energy is the temperature dependence of carrier concentration and mobility (see Fig.~\ref{FignmuvsT}). Therefore the coupling between spins is temperature-dependent. The corresponding $\Theta_{F}\simeq \Theta/{\bar x}$ amounts to approximately 755 K.

\begin{figure}[t]
 \includegraphics[width = 0.42\textwidth, bb = 0 20 590 530]
 {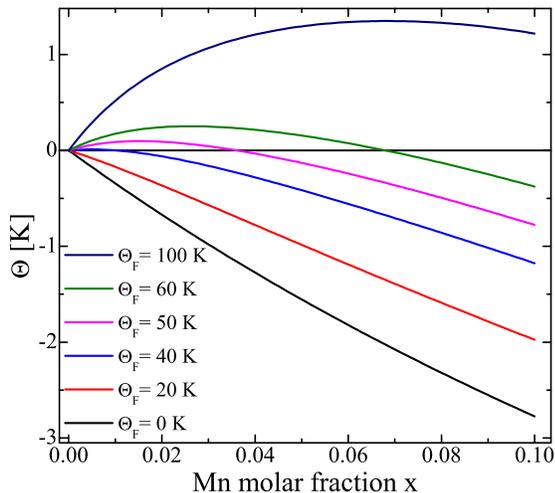}%
 \caption{\label{FigThetavsX} The calculated Curie-Weiss temperature as a function of the Mn content, $x$, for selected values of the characteristic ferromagnetic temperature $\Theta_{F}$, based on data for Cd$_{1\textrm{-}x}$Mn$_{x}$Te, a representative DMS with random distribution of magnetic impurities.}
\end{figure}

Let us state that in presence of both ferro- and antiferromagnetic interactions, the increase in $x$ may result in a sign change of Curie-Weiss temperature, $\Theta={\bar x}\Theta_{F}+\Theta_{AF}$($x$). The characteristic antiferromagnetic temperature, $\Theta_{AF}<0$, is a function of magnetic component concentration $x$. Its dependence on $x$ is known from experimental data for such diluted magnetic semiconductors as Zn$_{1-x}$Mn$_{x}$Te and Cd$_{1-x}$Mn$_{x}$Te \cite{Cibert08,Ferrand01}. Also, a dependence of ${\bar x}$ is known for such alloys. In already mentioned typical DMSs, the magnetic ion distribution is found to be random and uncorrelated. We can model qualitatively the $\Theta$ sign change using the dependencies of ${\bar x}$ and $\Theta_{AF}$ taken from \cite{Cibert08,Gaj94} for Cd$_{1-x}$Mn$_{x}$Te. The results are presented in Fig.$\;$\ref{FigThetavsX}, where predicted Curie-Weiss temperatures are plotted for various values of $\Theta_F$, which describes the total strength of long-range ferromagnetic interactions.
It is visible that in the presence of moderately strong ferromagnetic interactions we observe a sign change in $\Theta$. However, it is evident that this transition takes place for considerably higher concentrations of magnetic ions and the slope of Curie-Weiss temperature change is lower than that observed experimentally for our samples. This supports the picture of non-random distribution of magnetic ions, with significantly stronger antiferromagnetic contribution than expected from the presented model.

On the basis of the determined value of $J_{pd}$, we can make an attempt to predict the characteristic temperatures $\Theta_{F}$ for our samples with Mn content higher than $x=0.003$. For this purpose we take into account the low-temperature charge carrier concentrations and mobilities and we use the Eq.~\ref{Eq05} with an assumption that $J_{pd}$ is composition-independent. In this manner, for a sample with $x=0.014$ we obtain $\Theta_{F}=1465$ K; for $x=0.027$ we have $\Theta_{F}=165$ $K$ and for $x=0.042$ we get $\Theta_{F}=1040$ $K$. Note that if the distribution of Mn ions would be random in those samples, then the values of ${\bar x}\Theta_{F}$ would be high. Having the experimental values of Curie-Weiss temperatures, we can estimate the characteristic parameters $\Theta_{AF}$ for our samples from $\Theta={\bar x}\Theta_{F}+\Theta_{AF}$. As a consequence, we get the values of $\Theta_{AF}$ equal to -15.4$\;$K, -3.4$\;$K and -11.4$\;$K for the samples with $x=0.014$, $x=0.028$ and $x=0.042$, accordingly. The characteristic antiferromagnetic temperatures are quite high in magnitude (especially when compared, for example, with the values for Cd$_{1-x}$Mn$_{x}$Te \cite{Cibert08,Gaj94}). This fact supports the conclusion that the distribution of Mn ions in our samples is far from random and that the tendency towards creation of antiferromagnetically coupled clusters is promoted.

\section{Summary}

We explored the structural, electrical, and magnetic properties of the low-Mn-dilution limit of Zn$_{1\textrm{-}x}$Mn$_{x}$GeAs$_{2}$ alloys with 0$\,$$\leq$$\,$$x$$\,$$\leq$$\,$0.042. The XRD results indicate the appearance of chalcopyrite and disordered zincblende structure in our samples, strongly Mn content dependent.  \\ \indent The transport characterization shows that all our samples have high $p$ type conductivity with carrier concentration $n$$\,$$>$$\,$10$^{19}$$\;$cm$^{-3}$ and mobility $\mu$$\,$$<$$\,$50$\;$cm$^{2}$/(V$\cdot$s). The magnetoresistance in our samples shows large values, up to -50\%, for Zn$_{0.997}$Mn$_{0.003}$GeAs$_{2}$. The values are not correlated with magnetic properties of the alloy. Thus, the observed MR is not related to magnetic impurity scattering mechanism that would lead to MR proportional to the amount of magnetic impurities in the crystal. WL is known to be destroyed by the presence of magnetic impurities in the material. Therefore, the destruction of the WL by the presence of an increasing number of magnetic impurities in the crystals explains the observed MR in the Zn$_{1\textrm{-}x}$Mn$_{x}$GeAs$_{2}$ samples with $x$$\,$$>$$\,$0.003. \\ \indent The magnetic properties of the alloy indicate that in the case of the sample with very low Mn content, $x$$\,$$=$$\,$0.003, the majority of Mn ions are in a high-spin Mn$^{2+}$ charge state. For the samples with higher Mn content a large fraction of Mn ions forms antiferromagnetic pairs and/or clusters and stays in other, low moment charge states. The Curie-Weiss temperature, as determined from low-temperature magnetic susceptibility data, has small values $|$$\Theta$$|$$\,$$<$$\,$3$\;$K and changes sign from positive for $x$$\,$$=$$\,$0.003 into negative for higher Mn content. The Mn-hole exchange integral, estimated for the very diluted Zn$_{0.997}$Mn$_{0.003}$GeAs$_{2}$ sample is equal to $J_{pd}$$\,$$=$$\,$(0.75$\pm$0.09)$\;$eV indicating rather strong magnetic interactions in this material.

\section{Acknowledgments}

\noindent Scientific work was financed from funds for science in 2011-2014, under the project no.  N202 166840 granted by the National Center for Science of Poland. This work has been supported by the RFBR projects no. 12-03-31203 and 13-03-00125.

\noindent This work has been supported by the Polish Ministry of Science and Higher Education on a special purpose grant to fund the research and development activities and tasks associated with them, serving the development of young scientists and doctoral students.

\end{document}